\documentclass{jpsj-suppl}
\usepackage{txfonts} 
\usepackage{color}%
\title{Mean Field Voter Model of Election to the House of Representatives in Japan}

\author{Fumiaki \textsc{Sano}$^{1}$, Masato \textsc{Hisakado}$^{2}$, and Shintaro \textsc{Mori}$^{1}$}

\inst{$^{1}$Department of Physics, Faculty of Science, Kitasato University \\
Kitasato 1-15-1, Sagamihara, Kanagawa 252-0373, JAPAN \\
$^{2}$
Fintech Lab. LLC, Meguro, Tokyo 153-0051, JAPAN  
}

\email{mori@sci.kitasato-u.ac.jp}

\recdate{October 31, 2016}

\abst{ 
  In this study, we propose a mechanical model of a
  plurality election based on a mean
  field voter model.
  We assume that there are three candidates in each electoral
  district, i.e., one from the ruling party, one from the main opposition party, and
  one from other political parties. The voters are classified as fixed
  supporters and herding (floating) voters with ratios of $1-p$ and $p$,
  respectively. Fixed supporters make decisions based on their information
  and herding voters make the same choice as another randomly selected voter.
  The equilibrium vote-share probability density of herding voters
  follows a Dirichlet distribution. We estimate the composition of fixed
  supporters in each electoral district and $p$ using data
  from elections to the House of Representatives in Japan (43rd to 47th).
  The spatial inhomogeneity of fixed supporters explains the long-range
  spatial and temporal correlations. The estimated values of $p$ are
  close to the estimates obtained from a survey. 
}

\kword{correlation, Dirichlet distribution,
  mean field voter model, plurality election data}

\begin{document}
\maketitle

\section{Introduction}
Social phenomena are an active research field in econophysics
and socio-physics, and many studies have aimed to
deepen our understanding of
them\cite{Mantegna:2007, Pentland:2014, Ormerod:2012,Mori:2016}.
Opinion dynamics is a central research theme
and voter models have been studied extensively as a paradigm of opinion
dynamics\cite{Castellano:2009,Mori:2010,Hisakado:2010,Mori:2012}.
Recently, the validity of a model was tested
for describing the real opinion
dynamics in the U.S. presidential election\cite{Gracia:2014}, where
it was concluded that a noisy diffusive
process of opinions can reproduce the statistical features
of elections, i.e., the stationary vote-share distributions and
long-range spatial correlation, which decay logarithmically with
distance. The model is simple and
attractive in the domain of physics.

In the model, the decisions made by all the voters 
are described by an infection mechanism.
Two voters are selected randomly and one voter's choice is made the same as
another's choice, with some noise. 
The spatial inhomogeneity of the system is considered by using data
to define the initial conditions.
The noise level has to be fine tuned in order to simulate the statistical
properties of elections, particularly maintaining the initial spatial
inhomogeneity;
otherwise, the spatial pattern disappears or
the spatial correlation becomes too strong and the model
cannot reproduce the statistical properties of the election data.
However, it is necessary to avoid making subtle choices for the parameters 
 when modeling a social system.
If people interact with others and make decisions in a social system,
the process should be robust and stable. This
is the first problem that needs to be addressed.

The second problem is estimating the ratio of
floating voters who do not vote for a particular political party.
In the voter model described above,
the opinions of all the voters can be changed by 
interactions with other voters. However, this approach is too simple to
represent the situation in actual elections.
Some people prefer a certain political party and
it is reasonable to assume that their opinions will not be changed easily
 by interacting with others. 
 Nowadays, many people recently have become dissatisfied with
 the current status of the political
 system and they have no particular
 political party to vote for\cite{Killan:2012}.
They are called floating voters and it is considered that they have
a decisive role in the results of elections.
If we assume that social interactions play a
crucial role in their decisions 
\cite{Gracia:2014,Ormerod:2012,Araujo:2010}, then they should be studied
in the framework of a voter model.

In the present study, we propose a
mean field voter model to describe the dynamics of a plurality election in
Japan. In the model, voters are classified as fixed supporters
who have a preference for a specific political party and
floating voters whose decision depends on the choices of others.
The influence of other voters on floating voters is described
by the voter model mechanism. The vote share distribution of
floating voters follows a Dirichlet distribution and the system is stable.
We show how to decompose the vote share into votes
by fixed supporters and those by floating voters, and we estimate
the spatial inhomogeneity of the electoral system.
We study the fluctuations in the vote share of floating voters
and the ratio of floating voters is estimated.
We explain the spatial correlation and temporal
correlation in terms of the spatial inhomogeneity in the fixed
supporters.

\section{Mean Field Voter Model}

\begin{figure}[tbh]
\includegraphics[width=7cm]{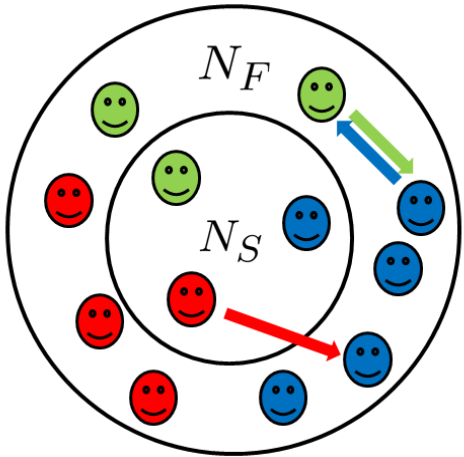}
\caption{Composition of voters.
    Voters are classified as
  fixed supporters in the inner circle 
  and floating voters in the area surrounded by inner and outer circles.
  The number of fixed (floating) voters is $N_{S} (N_{F})$.
  We assume $K=3$ political parties and 
  the voters are distinguished by painting in three color:
  red, blue and green for voter to political party 1,2 and 3, respectively.
  Fixed supporters decide based on their preferences
  and they are not affected by the choices of other voters.
  They only affect the choices of floating
  supporters. The  red arrow shows the influence
  from a fixed supporter of political party 1 to a
  floating supporter of political party 2.
  By contrast, the choices of floating supporters are influenced by
  those made by both fixed supporters and floating supporters.
  The blue (green) arrow shows the influence from
  a floating supporter of political party 2 (3)
  to a floating supporter of political party 3 (2). 
}
\label{fig1}
\end{figure}

There are $K$ political parties, $I$ electoral districts, and
$T$ elections. In each election, $K$ candidates fight for a single congress
seat in each electoral district.
We denote the political parties as $k\in \{1,2,\cdots,K\}$,
electoral districts as $i\in \{1,2,\cdots,I\}$, and
elections as $t\in \{1,2,\cdots,T\}$.
There are $N(t,i)$ votes in district $i$ for election $t$,
and $N(t,i,k)$ voters
vote for political party $k$.
$N(t,i)=\sum_{k}N(t,i,k)$ holds. 
$N(t,i,k)$ are classified as $N_{F}(t,i,k)$
floating (herding) voters and $N_{S}(t,i,k)$ fixed supporters who vote
for political party $k$ (Figure \ref{fig1}).
$N(t,i,k)=N_{F}(t,i,k)+N_{S}(t,i,k)$ and
we denote the total floating voters and supporters as $N_{F}(t,i)\equiv
\sum_{k}N_{F}(t,i,k)$
and $N_{S}(t,i)\equiv \sum_{k}N_{S}(t,i,k)$, respectively.
We write the ratio of floating voter as $p(t,i)$,
$p(t,i)\equiv N_{F}(t,i)/N(t,i)$.
We denote the vote share for political party $k$ as $Z(t,i,k)
\equiv N(t,i,k)/N(t,i)$. $Z(t,i,k)$ is then decomposed as
\[
Z(t,i,k)=(1-p(t,i))\cdot N_{S}(t,i,k)/N_{S}(t,i)+p(t,i)
\cdot N_{F}(t,i,k)/N_{F}(t,i).
\]
We introduce $\mu(t,i,k)\equiv N_{S}(t,i,k)/N_{S}(t,i)$, which represents
the vote share of political party $k$ among $N_{S}(t,i)$ fixed supporters.
The ratio $N_{F}(t,i,k)/N_{F}(t,i)$ is denoted as $X(t,i,k)$, which
represents the vote share of political party $k$ among
$N_{F}(t,i)$ floating voters.
\begin{equation}
Z(t,i,k)=(1-p(t,i))\cdot \mu(t,i,k)+p(t,i)\cdot X(t,i,k) \label{eq:decom}.
\end{equation}
We assume that the choices of fixed supporters do not change
and that the choices of floating voters are described by a voter
model. In each turn, a randomly selected floating voter
changes his choice to that 
of another randomly selected voter. The fixed supporters
 only affect the choices made by floating voters.
We also assume that
among $N_{S}$ fixed supporters, only 
$\phi, 0\le \phi \le N_{S}$ 
supporters can affect the choices of floating voters.
As shown in the following, 
$\phi$ controls the fluctuations in $X(t,i,k)$ from
$\mu(t,i,k)$.

At equilibrium, $K$ variables $X(t,i,k)$ with values in the standard unit
interval $(0,1)$ add up to 1, $\sum_{k}X(t,i,k)=1$, which constrains
the sample space of $K$ dependent variables to a $K-1$ dimensional
simplex. Thus, one variable $X(t,i,K)$ can always be omitted
due to $X(t,i,K)=1-\sum_{k=1}^{K-1}X(t,i,k)$.
We use the abbreviated form of
$\vec{X}(t,i)=(X(t,i,1),\cdots,X(t,i,K))$ and 
$\vec{\mu}(t,i)=(\mu(t,i,1),\cdots,\mu(t,i,K))$.
The probability density of $\vec{X}(t,i)$ is described
by a Dirichlet distribution,
which is given by Eq. (\ref{eq:Dirichlet}).
We employ the alternative parametrization of the Dirichlet
distribution introduced by Ferrari and Cribari-Neto\cite{Ferrari:2004}.
The derivation is given in Appendix B.
\begin{equation}
P(\vec{x}|\vec{\mu}(t,i),\phi)=\frac{1}{\mbox{B}(\phi \vec{\mu}(t,i))}
\prod_{k}x_{k}^{\mu(t,i,k)\phi-1}. \label{eq:Dirichlet}
\end{equation}
The denominator  
$\mbox{B}((\phi \vec{\mu}(t,i)))$ in eq.(\ref{eq:Dirichlet}) is the
multinomial beta function, which
serves as a normalization constant, and it is defined as
\[
\mbox{B}((\phi \vec{\mu}(t,i)))\equiv 
\frac{1}{\Gamma(\phi)}\prod_{k}\Gamma(\phi \mu(t,i,k)).
\]
$\Gamma(\cdot)$ is the gamma function defined as $\Gamma(x)
=\int_{0}^{\infty}t^{x-1}e^{-t}dt$.

We write the random variable $\vec{X}(t,i)$ obeys
the Dirichlet
distribution in the alternative parametrization $\vec{\mu}(t,i),\phi$
as 
\[
\vec{X}(t,i)\sim {\it D_{a}}(\vec{\mu}(t,i),\phi).
\]

Each component $X(t,i,k)$
of $\vec{X}(t,i)$ is marginally beta-distributed with
$\alpha=\phi \cdot \mu(t,i,k)$ and $\beta=\phi\cdot (1-\mu(t,i,k))$.
The expected values are defined as $\mbox{E}(\vec{X}(t,i))
=\vec{\mu}(t,i)$, the variances are
$\mbox{V}(X(t,i,k))=\mu(t,i,k)(1-\mu(t,i,k))/(\phi+1)$, and
the covariances are $\mbox{C}(X(t,i,k),X(t,i,l))=-\mu(t,i,k)\mu(t,i,l)
/(\phi+1)$. $\phi$ is a ``precision'' parameter to model
the dispersion of the variables.
In the context of the mean field voter model,
a small $\phi$ indicates that the influence is weak and
the fluctuations in $X(t,i,k)$ are large.
If $\phi$ is large, the influence is strong and
the fluctuations in $X(t,i,k)$ from $\mu(t,i,k)$
are suppressed, and $X(t,i,k)$
almost coincides with $\mu(t,i,k)$.

\section{Data Analysis}

\begin{table}[tbh]
  \caption{House of Representatives (general) Election in Japan.
    We use data from the 43rd to 47th plurality
  elections.
  }
\label{table:1}
\begin{tabular}{|l|l|ccccc|}
\hline
$t$ & date & Districts & Regions & Voters & Votes & Ruling Political Party    \\ 
\hline  
1 & 2003/11/9 & 300  & 3345  & $1.02 \times 10^{8}$ & $6.12\times 10^{7}$ & LDP     \\
2 & 2005/9/11 & 300  & 2534  & $1.03\times 10^{8}$ & $6.95\times 10^{7}$  & LDP  \\
3 & 2009/8/30 & 300  & 2037 & $1.04\times 10^{8}$ & $7.20\times 10^{7}$  & LDP \\
4 & 2012/12/16 & 300  & 1994  & $1.04\times 10^{8}$ & $6.17\times 10^{7}$ & DPJ  \\
5 & 2014/12/14 & 295  & 1983  & $1.04\times 10^{8}$ & $5.47\times 10^{7}$ & LDP  \\
\hline
\end{tabular}
\end{table}

We employ data from elections to the
House of Representatives (general election) in Japan.
This is a plurality election and there are about 300 electoral
districts, where several candidates compete for a single congress seat.
We use data  of $T=5$ elections 
from the 43rd (2003) to 47th (2014) elections, which we label as 
$t=1,2,3,4,$ and $5$ in Table \ref{table:1}.
During this period, the Liberal Democratic party of Japan (LDP)
and the Democratic party of Japan (DPJ) were elected as
the ruling party in $t=1,2,3$ and $5$ elections and $t=4$ election,
respectively. 
As there is only one congress seat in each electoral district,
there is at most one candidate from each of these two political parties.
We label the candidates from LDP as $k=1$ and  from DPJ
as $k=2$.
There might be several candidates from other political parties (OPP),
we combine them together and treat them as one candidate.
We label them as  $k=3$.
By this treatment of several candidates from OPP,
there are at
most three candidates in each district.
Each district is divided into several regions where
the results of the elections were recorded.
There are 2000 to 3000 regions.

\begin{figure}[tbh] 
\includegraphics[width=15cm]{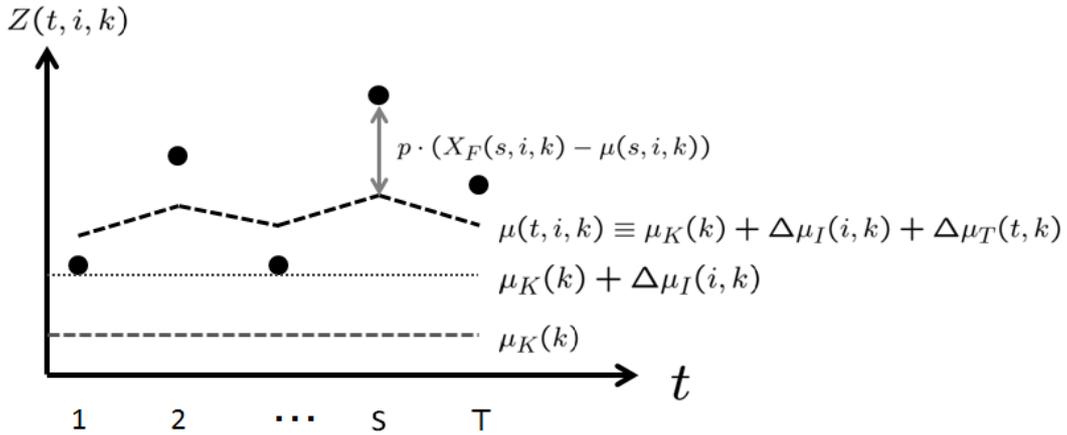}
\caption{Decomposition of $Z(t,i,k)$.
  $Z(t,i,k)$ are decomposed as $Z(t,i,k)=(1-p(t,i))
  \mu(t,i,k)+p(t,i)\cdot X(t,i,k)$ in Eq. (\ref{eq:decom}).
  $\mu_{K}(k)$ is the fitness of political party $k$ and
  $\mu_{K}(k)+\Delta \mu_{I}(i,k)$ is the fitness of a candidate from political
  party $k$ in region $i$. $\Delta \mu_{T}(t,k)$ represents the trend for political
  party $k$ in election $t$.}
\label{fig2}
\end{figure}

We use data from the regions where  the above  three candidates
fought for a single seat five times in a row due to the
necessity to infer $\mu(t,i,k)$ in the mean field voter model.
There are $I=488$ regions, which we label as $i=1,2,\cdots,I$.
We denote the vote share for candidate $k$ in region $i$ in election $t$
as $Z(t,i,k)$. $\sum_{k}Z(t,i,k)=1$ holds in region $i$ in election
$t$. In order to decompose $Z(t,i,k)$ as described in
Eq. (\ref{eq:decom}), we write $Z(t,i,k)$ as
\begin{eqnarray}
Z(t,i,k)&=& \mu(t,i,k)+(Z(t,i,k)-\mu(t,i,k)), \nonumber \\
\mu(t,i,k)&\equiv& \mu_{K}(k)+\Delta \mu_{T}(t,k)+\Delta \mu_{I}(i,k), \nonumber \\
\mu_{K}(k)&\equiv &\frac{1}{IT}\sum_{t}\sum_{i}Z(t,i,k), \nonumber \\
\Delta \mu_{T}(t,k)&\equiv &\frac{1}{I}\sum_{i}Z(t,i,k)-\mu_{K}(k), \nonumber \\
\Delta \mu_{I}(i,k)&\equiv &\frac{1}{T}\sum_{t}Z(t,i,k)-\mu_{K}(k). \nonumber 
\end{eqnarray}
$\mu_{K},\Delta \mu_{T},\Delta \mu_{I}$
shows the different types of 
fitness values for the politicians from
political party $k$. Please refer to Figure \ref{fig2}.
$\mu_{k}$ shows the overall fitness, and 
$\Delta \mu_{T}$ and $\Delta \mu_{I}$ show the temporal and regional
deviation from $\mu_{K}$, respectively. $\mu(t,i,k)$ shows the fitness
of a candidate from political party $k$ in region $i$ in
election $t$.

\begin{table}[t]
\caption{Symbol legend}
\label{tab:notation}
\begin{tabular}{l l}
\hline
$K,k$ & Number of political parties and their index  \\  
$T,t$ & Number of elections and time variable  \\
$I,i$ & Number of regions and their index \\
$N(t,i)$ & Number of votes in region $i$ for  election $t$   \\ 
$N(t,i,k)$ & $N(t,i)$ for political party $k$   \\
$Z(t,i,k)\equiv N(t,i,k)/N(t,i)$ & Votes share for political party $k$ in region $i$ for election $t$   \\
\hline
$N_{F}(t,i),N_{S}(t,i)$ & Number of floating voters and fixed supporters in region $i$, election $t$   \\ 
$p(t,i)=N_{F}(t,i)/N(t,i)$ & Ratio of floating voter  \\ 
$N_{S}(t,i,k)$ &  $N_{S}(t,i)$ for political party $k$ \\ 
$\mu(t,i,k)\equiv N_{S}(t,i,k)/N_{S}(t,i)$ &
Vote share of
political party $k$ among $N_{S}(t,i)$ fixed supporters\\ 
$\mu_{K}(k)\equiv \sum_{i,t}Z(t,i,k)/IT$ & Overall fitness of political party $k$  \\ 
$\Delta \mu_{T}(t,k)\equiv \sum_{i}Z(t,i,k)-\mu_{K}(k)$ &
Temporal deviation of fitness from $\mu_{K}$,\\ 
$\Delta \mu_{I}(i,k)\equiv \sum_{t}Z(t,i,k)-\mu_{K}(k)$ &
Regional deviation of fitness from $\mu_{K}$,\\ 
$N_{F}(t,i,k)$ &  $N_{F}(t,i)$ for political party $k$ \\ 
$X(t,i,k)\equiv N_{F}(t,i,k)/N_{F}(t,i)$ & Vote share of
political party $k$ among $N_{F}(t,i)$ floating voters \\
$\phi$  & Number of fixed supporters who can affect the choices of
floating voters.
\\
\hline
\end{tabular}

\end{table}

We summarize the notations in Table \ref{tab:notation}.
The decomposition is similar to ANOVA (analysis of variance ) in statistics.
There are three factors, $k,i$ and $t$ and we assume there is no
interaction effect among them and $\mu(t,i,k)$ are estimated as the sum
of the three factors, $\mu_{K}(k),\Delta \mu_{T}(t,k)$ and $\Delta \mu_{I}(i,k)$. 
In the next section, we use the maximum likelihood principle
to estimate the model parameters. Based on the results, we test
 the validity of the mean field voter model. 

\section{Results}
We assume that the floating voter ratio $p(t,i)$ does not depend on region
$i$ and write it as $p(t)$. In the mean field voter model,
$\vec{Z}(t,i)$ are decomposed as,
\begin{eqnarray}
\vec{Z}(i,t)&=&(1-p(t))\cdot \vec{\mu}(t,i)+p(t)\cdot \vec{X}(t,i), \nonumber \\
\vec{X}(t,i)&\sim & {\it D_{a}}(\vec{\mu}(t,i),\phi).
\end{eqnarray}
There are $T+1$ parameters $(\phi,p_{1},p_{2},\cdots,p_{T})$ in
the model, which we estimate using the maximum likelihood principle.
When we apply the maximum likelihood principle,
some $X(i,t,k)$ becomes negative for small values of $p$. In this case,
we assume that $p=1$. $\phi$ is estimated as $10.8\pm 0.2 \mbox{(SE)}$.
Table \ref{tab:2} shows the results for
$p_{1},\cdots,p_{T}$ in the first row.
We also show the number of cases
where we employed $p=1$ in the second row.
For about one-third of $I=488$ cases, at least one of $X(t,i,k)$ for
each $i$ becomes negative by the choice  $p(t)$ in election $t$.
Thus, the estimate of $\mu(t,i,k)$ is not appropriate or the assumption
that $p(t,i)$ is independent of $i$ is too crude.

\begin{table}[b]
  \caption{Maximum likelihood estimate for $p(t)$.
  The first row shows the results for $p(t)$
  and the second row shows the number of cases 
  where we assume that $p=1$ among $I=488$.
  The third row shows the results of a poll by Japan
  Broadcasting Corporation (NHK).}
\label{tab:2}
\begin{tabular}{l|ccccc}
\hline
$t$ & 1 & 2 & 3 & 4 & 5    \\ 
\hline  
$p(t)$ & $0.3666\pm 0.0009$ & $0.3234\pm 0.0005$ & $0.3201\pm 0.0007$ & $0.3551 \pm 0.0008$ &  $0.3314 \pm 0.0005$ \\
$p=1$ & 146 & 124 & 132 & 78 & 142 \\
NHK poll & $0.387$ & $0.256$  &  $0.295$ & $0.335$
& $0.263$ \\
\hline
\end{tabular}
\end{table}

We also show the estimates obtained by
Japan Broadcasting Corporation (NHK) in the third row,
which are based on a public poll of more
than 1000 people\cite{NHK}.
According to the poll, 25\% to  39\%
are estimated as floating voters.
The estimate obtained by the mean field voter model is in the
same order. 

\subsection{$\mu$ vs $Z$ and standardized $X_{R}$}

\begin{figure}[tbh] 
\includegraphics[width=8cm]{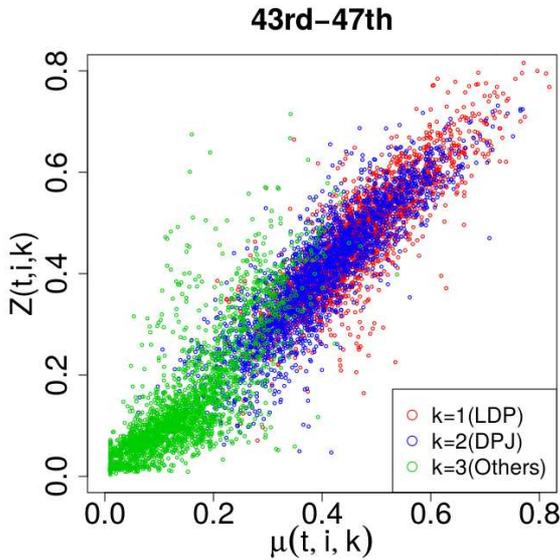}
\caption{Scatter plot of $\vec{\mu}(t,i)$ vs $\vec{Z}(t,i)$ for all
  data $t=1,2,3,4,5$. The red symbol denotes the plots for $k=1$, the 
  blue symbol represents those for $k=2$, and the green symbol indicates those
  for $k=3$.}
\label{fig3}
\end{figure}

We show the scatter plots for $\mu(t,i,k)$ and $Z(t,i,k)$
in Figure\ref{fig3}.
If the decomposition in Eq. (\ref{eq:decom}) is correct, then 
$Z(t,i,k)$ is distributed around $\mu(t,i,k)$. As shown in
Figure \ref{fig3}, $Z(t,i,k)$ for $k=1$ and $k=2$ are 
distributed almost symmetrically around the diagonal. However,
the distribution of $Z(t,i,3)$ is not symmetric around the diagonal,
so the decomposition might not be good for the
candidates with $k=3$.

\begin{figure}[htbp]
\begin{tabular}{ccc}
\includegraphics[width=4.5cm]{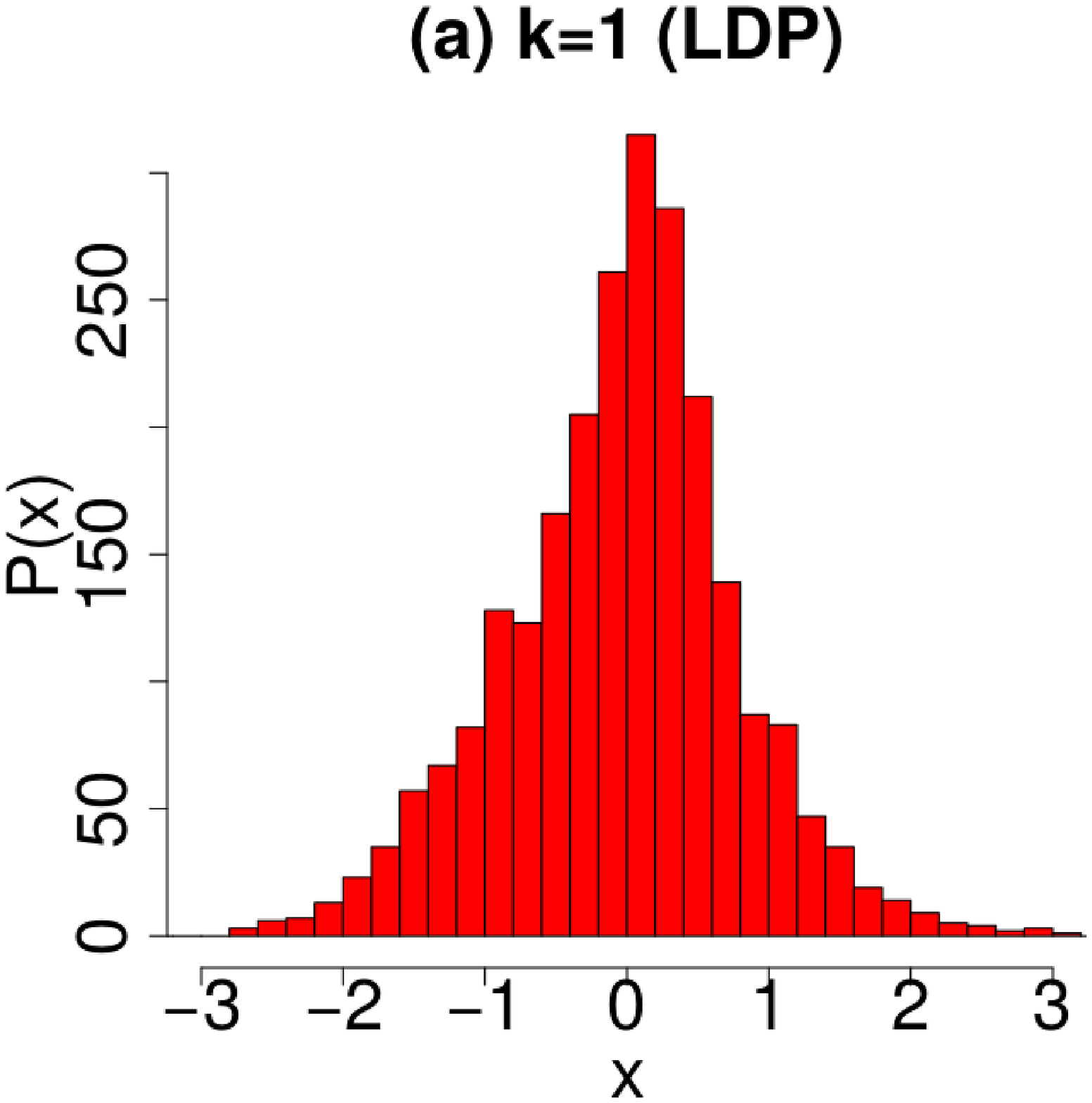} &
\includegraphics[width=4.5cm]{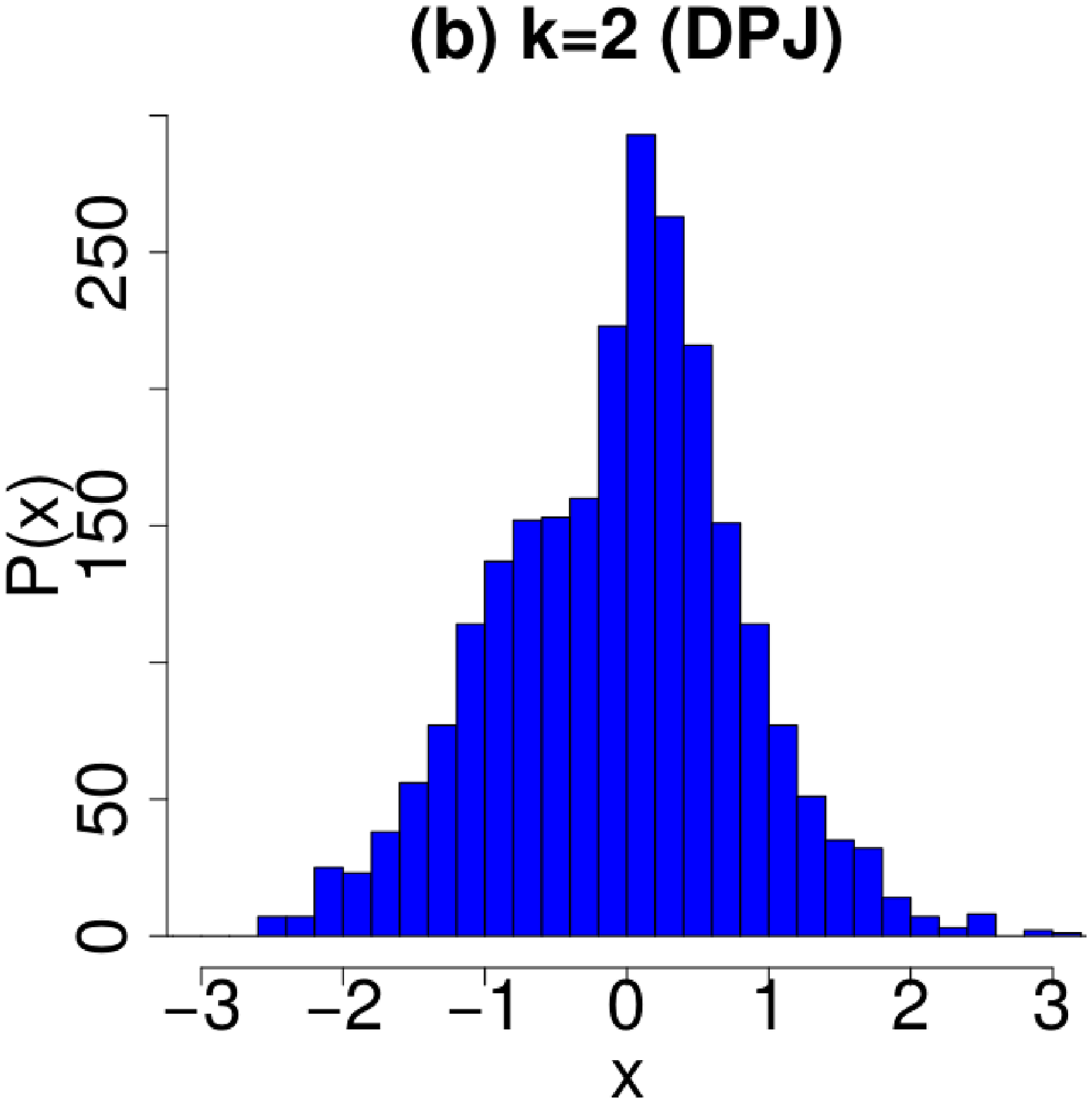} &
\includegraphics[width=4.5cm]{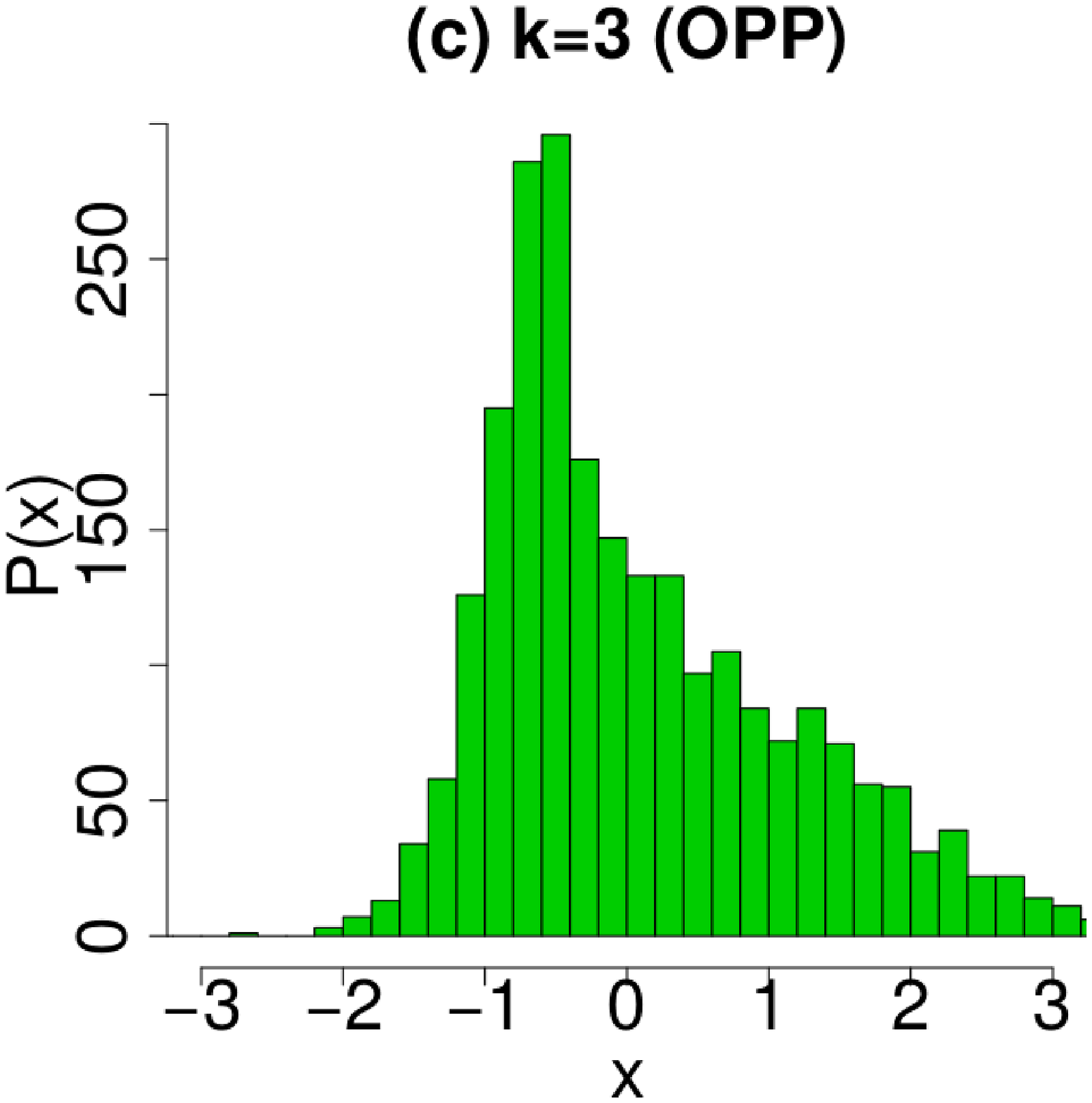} 
\end{tabular}
\caption{Probability density functions for
  renormalized $X_{R}(t,i,k)$ in Eq. (\ref{eq:renorm}), 
(a) $k=1$ (LDP), (b) $k=2$(DPJ) and (c) $k=3$ (other).  
The variances are estimated as $0.68,0.74$, and $1.59$ for $k=1,2$,
and $3$, respectively.  
}
\label{fig4}
\end{figure}	
In order to check the validity of the proposed model,
we estimate the standardized
$X(t,i,k)$, which is denoted as $X_{R}(t,i,k)$. If the model is correct and
$\vec{X}(t,i)\sim {\it D_{a}}(\vec{\mu}(t,i),\phi)$ holds, then 
V$(X(t,i,k))=\mu(t,i,k)(1-\mu(t,i,k)/(\phi+1)$.
$X_{R}(t,i,k)$ is defined as
\begin{equation}
  X_{R}(t,i,k)\equiv
  \sqrt{\phi+1} \frac{(X(t,i,k)-\mu(t,i,k))}{\sqrt{\mu(t,i,k)(1-\mu(t,i,k))}}
  \label{eq:renorm}.
\end{equation}
Figure \ref{fig4} shows the probability densities
for $X_{R}(t,i,k)$ with $k=1,2,3$, respectively.
The variances in $X_{R}(t,i,k)$ are estimated as $0.68,0.74$, and $1.59$
for $k=1,2,3$, respectively. As shown in
Figure \ref{fig4},
the distribution with $k=3$ is wider than those with $k=1,2$.
$X_{R}$ does not follow a normal distribution, so the asymmetrical nature
of the distribution is not crucial. However, the variance with
$k=3$ is about the twice that with $k=1,2$, which
suggests that the decomposition in Eq. (\ref{eq:decom}) 
is not good. We discuss improvements to the proposed model in the conclusion.
 
\subsection{Spatial and temporal correlations}

\begin{figure}[htbp]
\begin{tabular}{ccc}
\includegraphics[width=5cm]{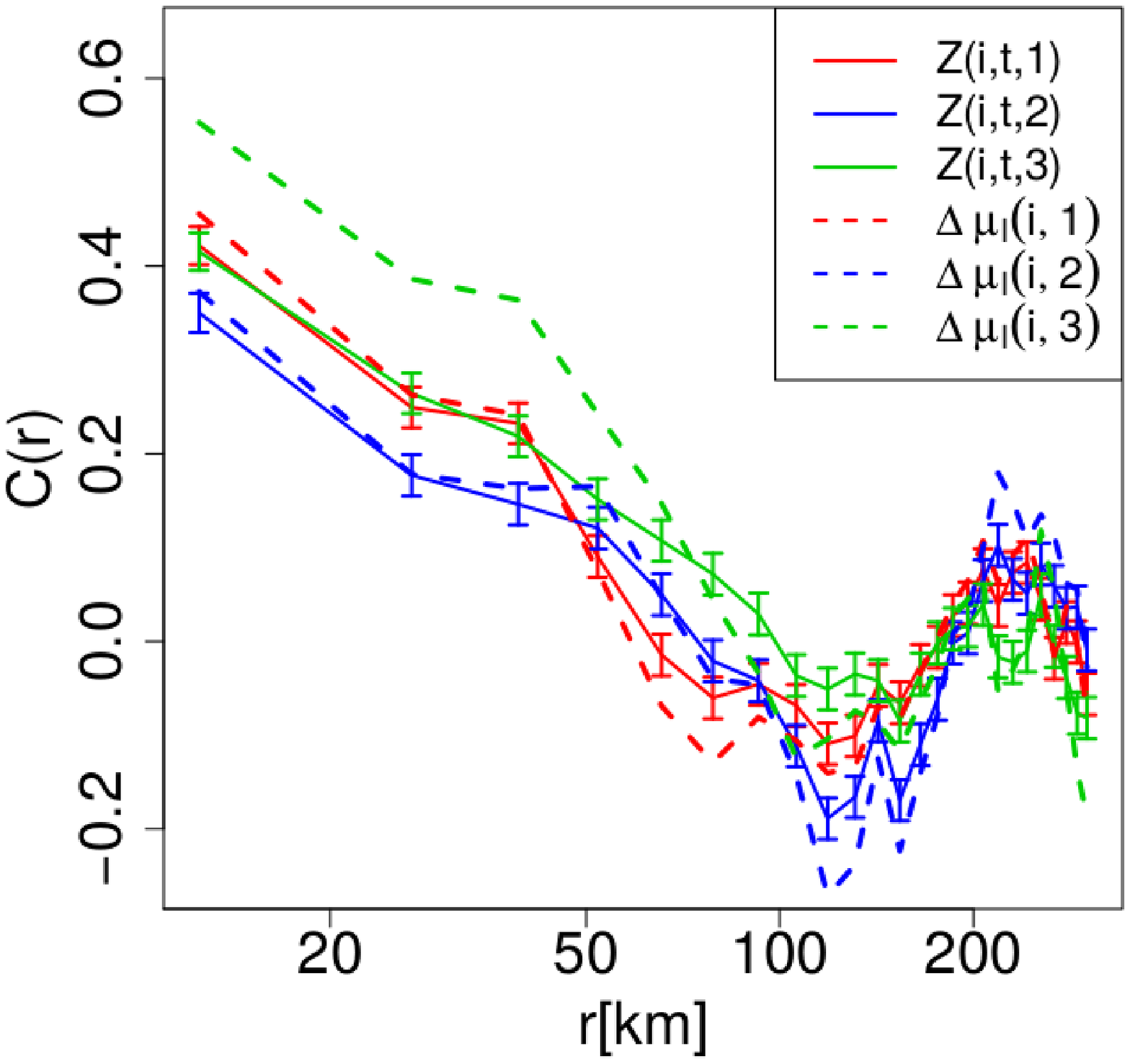} &
\includegraphics[width=5cm]{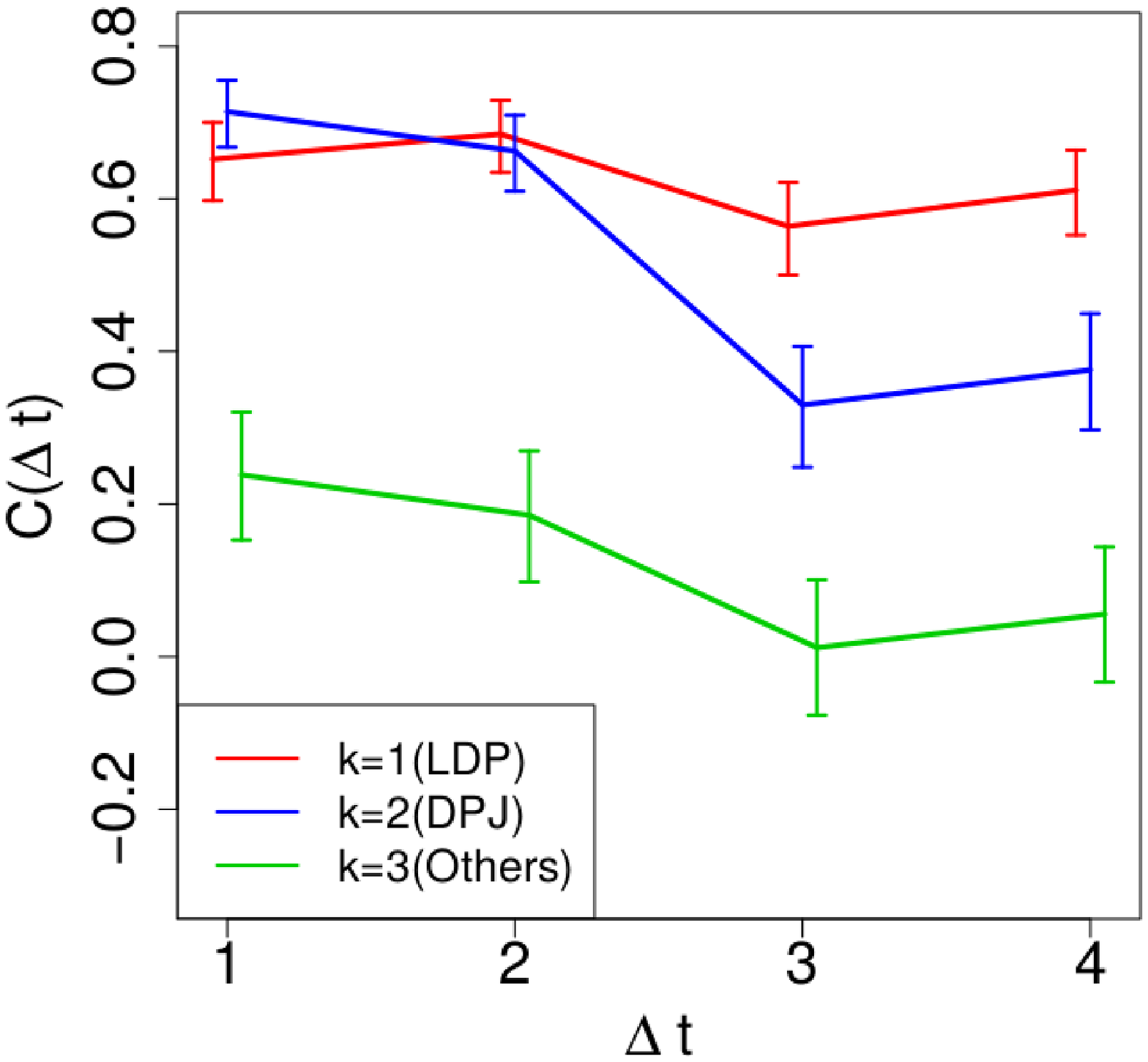} &
\includegraphics[width=5cm]{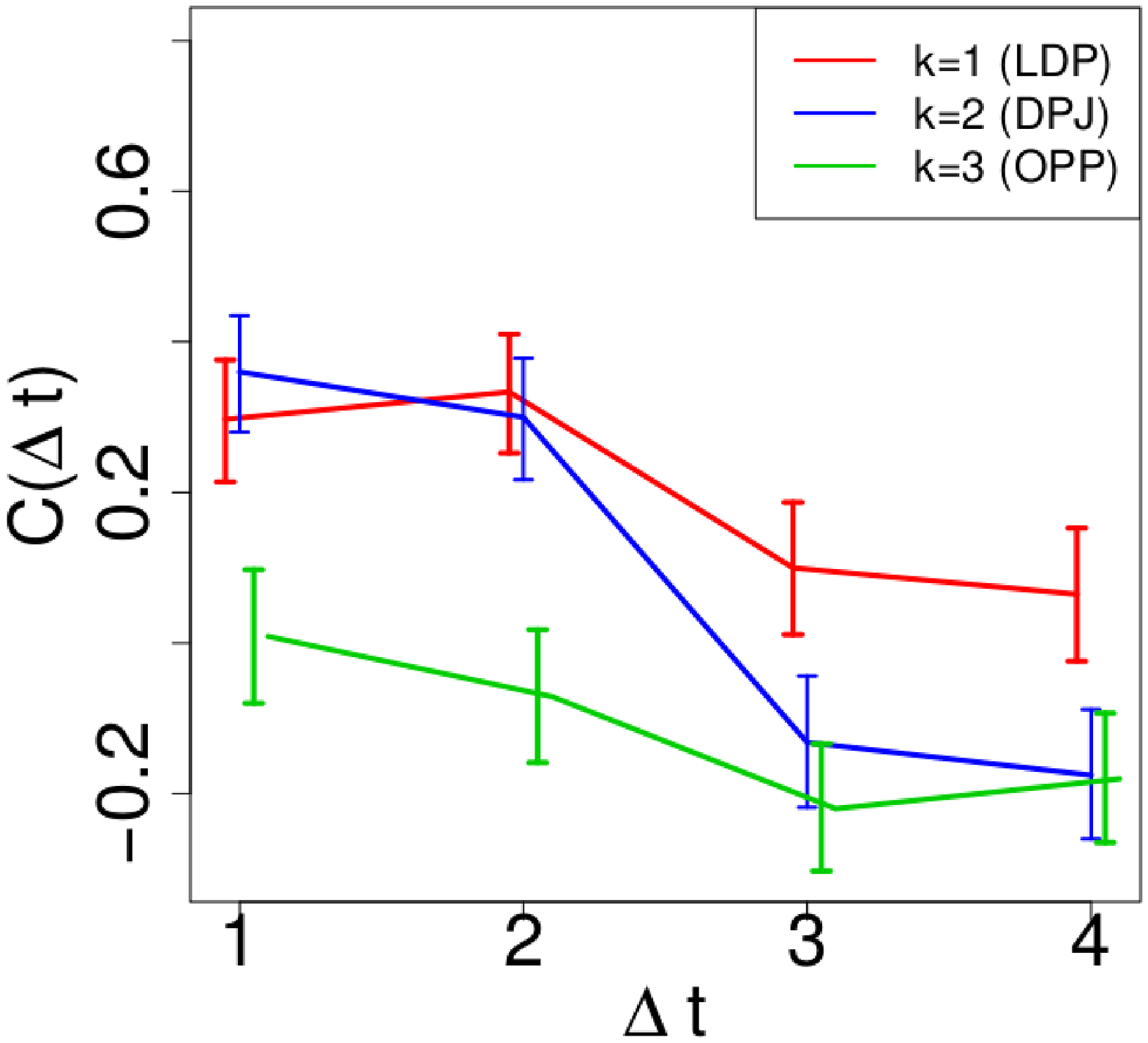} 
\end{tabular}
\caption{
(a) Spatial vote-share correlations and $\mu$ correlations
as a function of distance.
The solid line shows the spatial correlation for the vote share $Z(t,i,k)$
and the dashed lines indicate the
 spatial correlations for $\Delta \mu_{I}(i,k)$.
  (b) Temporal vote-share correlation for $Z(t,i,k) $ and (c) that 
  for $X(t,i,k)$ as a function of time.}
\label{fig5}
\end{figure}

We estimate the spatial and temporal correlations. 
The spatial correlation function is computed as
\begin{eqnarray}
C(r,t)&=&(\left<Z(t,i,k)Z(t,j,k)\right>|_{r(i,j)=r}-\mu_{T}(t,k)^{2})/\mbox{V}(\mu(t,k)),
  \nonumber \\
C(r)&=&\frac{1}{T}\sum_{t=1}^{T}C(r,t,k),
\nonumber \\
\mu_{T}(t,k)&\equiv& \sum_{i}Z(t,i,k)/I\,\,\, , \,\,\,  
\mbox{V}(\mu(t,k))\equiv \sum_{i}(Z(t,i,k)-\mu_{T}(t,k))^{2}/I
\end{eqnarray}
The first term $\left<Z(t,i,k)Z(t,j,k)\right>|_{r(i,j)=r}$
is averaged over pairs of regions separated by a distance $r$.
The spatial correlation decays logarithmically with 
geographical distance [Figure \ref{fig5}(a)], as reported in 
previous studies.
 The logarithmic decay of the spatial correlations is considered
 generic for the fluctuations in electoral dynamics\cite{Gracia:2014}.
 In the figure, we also plot the correlation functions for $\Delta \mu_{I}(i,k)$,
 which clearly almost coincide with $C(r)$ for $Z(t,i,k)$.
 This suggests that the physical description of the spatial correlation
 is spatial inhomogeneity in $\Delta \mu_{I}$. Every region has certain
 characteristics, such as the type of the region and its historical nature.
 These characteristic do not change rapidly even if some
people move between regions. If all the voters are floating
 voters\cite{Gracia:2014}, the diffusion and noise should be controlled
 to maintain the inhomogeneity in $\Delta \mu_{I}$. In our model, the voters are classified
 as fixed supporters and floating voters, and there is no movement
 of people among different regions, so it is not necessary to consider
 this issue.
 
 Figure \ref{fig5}(b) shows the results for the temporal correlation,
 $C(\Delta t)$, which is computed as
\begin{equation}
  C(\Delta t)=\mbox{Cov}(Z(1,i,k),Z(1+\Delta t,i,k))/\sqrt{\mbox{V}(Z(1,i,k))\mbox{V}(Z(1+\Delta t,i,k))}
\label{eq:Ct}.
\end{equation}
Here $\mbox{Cov}(A(i),B(i))$ is defined as $\sum_{i}A(i)B(i)/I
-\sum_{i}A(i)/I\cdot \sum_{j}B(j)/I$.
$C(\Delta t)$ does not decay with $\Delta t$, which can be explained
by the decomposition in Eq. (\ref{eq:decom}).
The covariance of $Z(1,i,k)$ and $Z(1+\Delta t,i,k)$ is decomposed into 
the covariance of $\mu(t,i,k)$ and $\mu(t+\Delta,i,k)$, and the covariance
of $X(t,i,k)$ and $X(t+\Delta,i,k)$.
\begin{eqnarray}
\mbox{Cov}(Z(t,i,k),Z(t+\Delta t,i,k))&=&
(1-p(1))(1-p(1+\Delta t))\mbox{Cov}(\mu(1,i,k),\mu(1+\Delta t,i,k))
\nonumber \\
&+&p(1)p(1+\Delta t)\mbox{Cov}(X(t,i,k),X(t,i,t+\Delta t))\nonumber .
\end{eqnarray}
We assume that $\mu(t,i,k)$ and $X(t,i,k)$ are independent from each other.
 If we further assume that the correlation between $X(1,i,k)$ and $X(1+\Delta t,i,k)$ 
 decays with $\Delta t$, as $\Delta \mu_{K}(k)$ and $\Delta \mu_{T}(t,k)$ do not
 depend on $i$, we have
\[
\mbox{Cov}(\mu(1,i,k),\mu(1+\Delta t,i,k))=\mbox{V}(\Delta \mu_{I}(i,k)).
\]
Then, the temporal correlation $C(\Delta t)$ for a large value of $\Delta t$ is
approximately expressed as
\begin{equation}
C(\Delta t)=(1-p(1))(1-p(1+\Delta t))\cdot \frac{\mbox{V}(\Delta \mu_{I}(i,k))}{
\sqrt{\mbox{V}(Z(1,i,k))\mbox{V}(Z(1+\Delta t,i,k))}}\label{eq:temp},
\end{equation}
which suggests that the physical origin of the temporal correlation
is also the spatial inhomogeneity of $Z_{I}$.

We also check the decomposition in Eq. (\ref{eq:decom}) by studying
the temporal correlation of $X(t,i,k)$.
Figure \ref{fig5}(c) shows it, 
which is defined by replacing $Z(t,i,k)$ with $X(t,i,k)$ in eq.
(\ref{eq:Ct}).
$C(\Delta t)$ decays with $\Delta t$ and it almost vanishes for $\Delta t=2$.
However, $C(\Delta t)$ becomes negative for $\Delta t\ge 3$ and $k=2,3$.
This suggests that a more subtle decomposition of $Z(t,i,k)$
should be performed.

\section{Conclusions}
In this study, we proposed a mean field voter model for a plurality election.
We assumed that voters are classified as fixed supporters and
floating voters, where the behavior of the latter can be described by the voter
model's infection mechanism. We decomposed the vote share $\vec{Z}$
into that from the fixed supporters $\vec{\mu}$
and that from the
floating voters $\vec{X}$, $\vec{Z}=(1-p)\cdot \vec{\mu}
+p\cdot \vec{X}$. 
$\vec{\mu}$ has three factors $\vec{\mu}_{K},\Delta \vec{\mu}_{T}$ and
$\Delta \vec{\mu}_{I}$.
$\vec{X}$ follows a Dirichlet
distribution with the parameters $\phi,\vec{\mu}$.
We estimated the model parameter $p,\phi,\vec{\mu}$
, where we used electoral data from the House of Representatives
elections in Japan during 2003--2014.
There we assume that $p$ does not depend on the region
and there is no interaction effect among the three
factors in $\vec{\mu}$. 
 Using the estimated parameters,
we decomposed $\vec{Z}$ and
checked the validity of the proposed model.

\begin{itemize}
\item The variances in the standardized $\vec{X}_{R}$ are $0.68,0.74$,
  and $1.59$ for $k=1,2,3$, respectively (Figure \ref{fig4}).
  
\item The spatial correlation functions for $\vec{Z}$ are described
  by those for $\Delta \mu_{I}(i,k)$ (Figure \ref{fig5}(a)).

\item The temporal correlation function for $\vec{Z}$ is explained 
  by the variance in $\Delta \mu_{I}(i,k)$ (Figure \ref{fig5}(b) and
  Eq. (\ref{eq:temp})).
\end{itemize}  
These results show that the decomposition of
$\vec{Z}=(1-p)\vec{\mu}+p\vec{X}$ is effective for simulating
the statistical nature of the electoral data. However,
the decomposition should be performed carefully to verify the validity of
the mean field voter model.
We estimated $\vec{\mu}$ using some average values of $\vec{Z}$,
which should be replaced
with the maximum likelihood estimates. In addition, the assumption of
the independence of 
$p(t,i)$ on $i$ is also excessively crude and it is necessary to consider
their regional dependencies. These issues should be addressed in future
studies.

\appendix
\section{Election Data}
\begin{figure}[tbh] 
\begin{tabular}{cc}
\includegraphics[width=7cm]{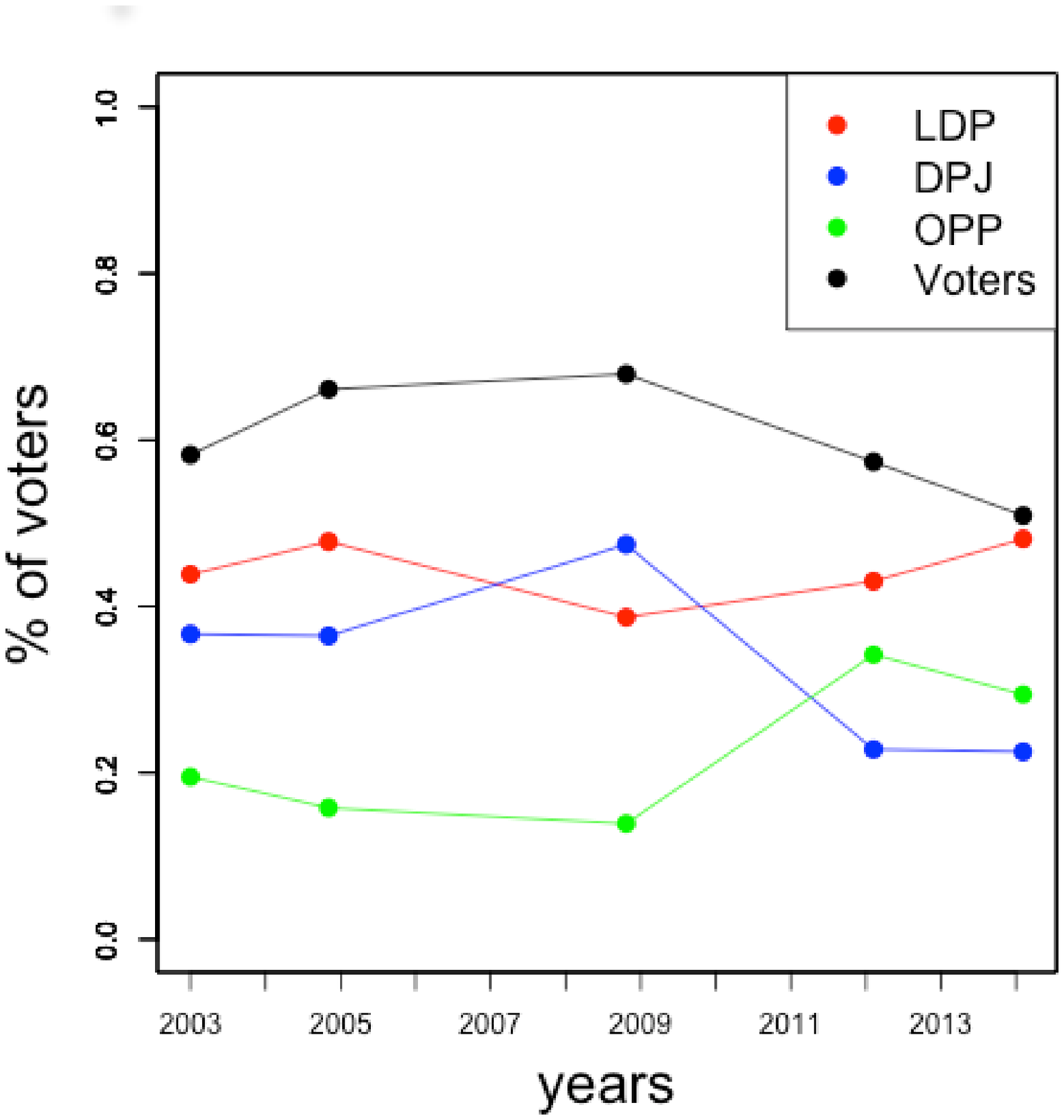} &
\includegraphics[width=7cm]{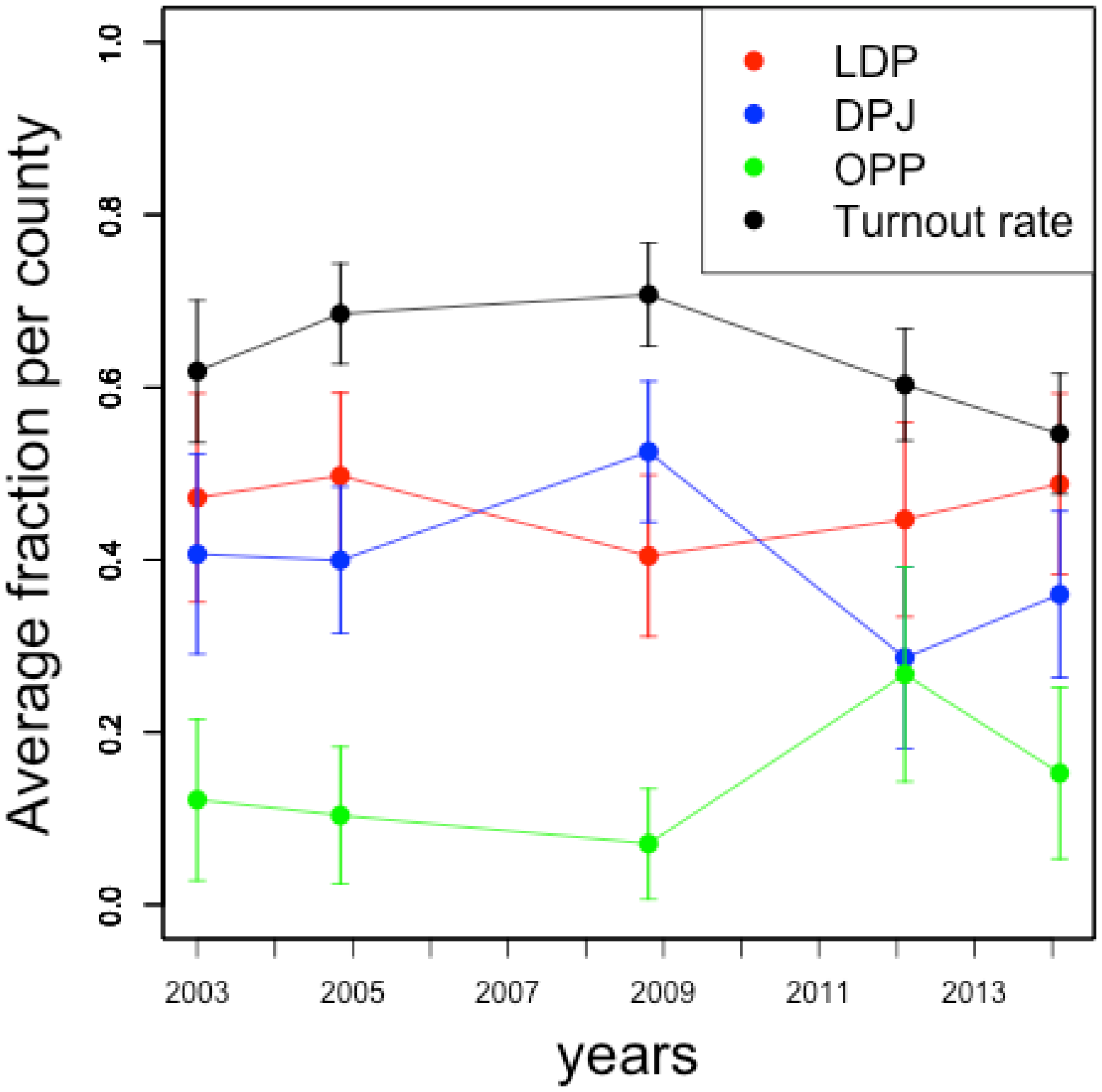} 
\end{tabular}
\caption{Japanese general election results. 
  (a) Global trends in the absolute values of different quantities, such as the
  turnout, and the votes for LDP, DPJ, and OPP. (b) Changes in the
  shares associated with the turnout and votes for different parties in
  $I=488$ regions where more than three candidates fought for a single seat
  five times in row.
}
\label{fig6}
\end{figure}

We provide further information regarding the data discussed in the
main text. We used the results of House of Representatives
elections from 2003 (43rd) to 2014 (47th), which were aggregated into
electoral districts\cite{Data}.
In Figure\ref{fig6}(a), we present the global features of the election data, i.e.,
the turnout as well as votes for the LDP, DPJ, and OPP.
In Figure \ref{fig6}(b), we show the changes in the
shares associated with the turnout and the votes for different parties
in the regions considered in the main text. There are $I=488$ regions.
The shares were computed region-by-region, and we then extracted the
averages and standard deviations.

\begin{figure}[htbp]
\includegraphics[width=15cm]{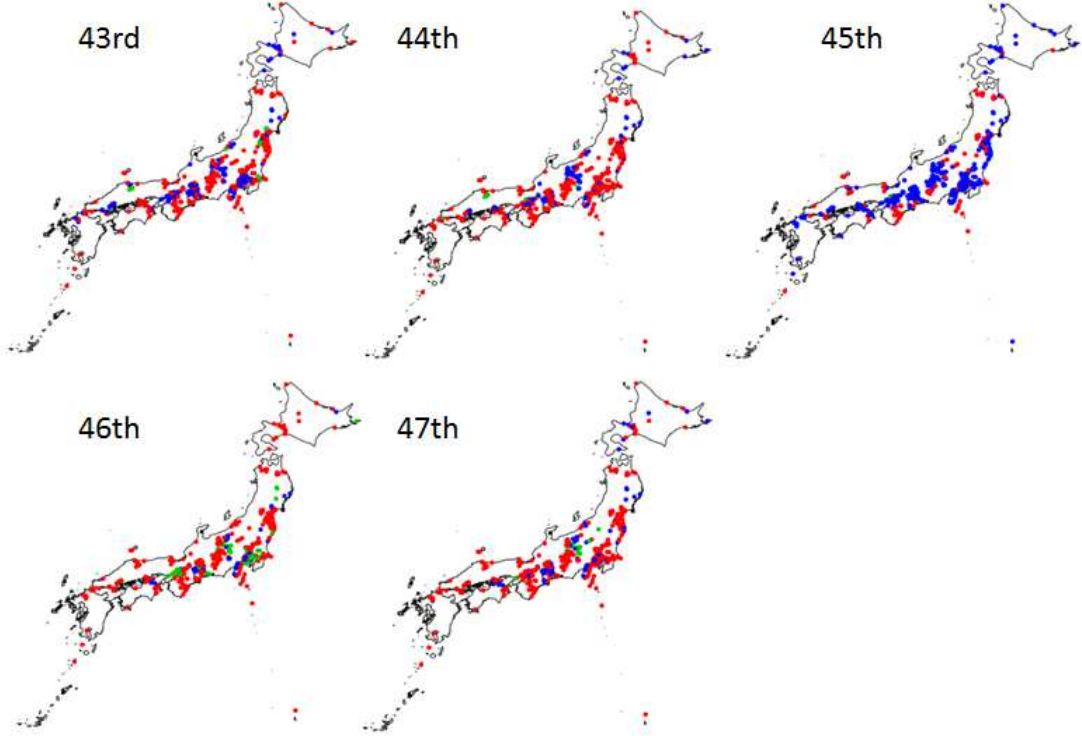} 
\caption{Schematic representation of the election results for the
  House of Representatives in 2003--2014.
  The winning political party is shown for $I=488$
  regions where candidates from more than three political parties fought
  four a single congress seat in all elections. The political parties
  are shown in red (LDP), blue (DPJ), and green (OPP).
}
\label{fig7}
\end{figure}

Figure \ref{fig7} shows the
winning political party in $I=488$ electoral regions during five general
elections. 

\section{$\vec{X}\sim {\it D_{a}}(\vec{\mu},\phi)$}
We derive the Dirichlet distribution for $\vec{X}$.
There are $N_{F}$ floating voters and $N_{i}$ voters for political party
$i$. $N_{F}=\sum_{i}N_{i}$. We write the probability function as 
$\mbox{Pr}(\vec{N}=\vec{n})=P(\vec{n})$.
The proportion of fixed supporters $\vec{\mu}$ affects
the choices of the floating
voters because there are $\phi \vec{\mu}$ voters. 
The probability that a randomly selected voter is a floating
voter with choice $i$ is $n_{i}/N_{F}$. The probability that
another randomly chosen voter makes the choice $j$ is
$(\phi \mu_{j}+n_{j})/(\phi+N_{F}-1)$.
The transition probability from state $(n_{i},n_{j}) \to (n_{i}-1,n_{j}+1)$
for any pair $(i,j), 1\le i\neq j\le K$
is given as
\[
\mbox{Pr}((n_{i},n_{j})\to (n_{i}-1,n_{j}+1))=\frac{n_{i}}{N_{F}}\cdot
\frac{\phi \mu_{j}+n_{j}}{N_{F}+\phi-1}.
\]
Similarly, the transition probability from state
$(n_{i}-1,n_{j}+1))\to (n_{i},n_{j})$ is given as
\[
\mbox{Pr}((n_{i}-1,n_{j}+1)\to (n_{i},n_{j}))=\frac{n_{j}+1}{N_{F}}\cdot
\frac{\phi \mu_{i}+n_{i}-1}{N_{F}+\phi-1}.
\]
The detailed balance condition for the stationary $P(\vec{n})$
 can be solved easily and we obtain
\[
P(\vec{n})=\frac{N_{F}!}{\prod_{j}n_{j}!}
\frac{\prod_{j}(\phi \mu_{j})^{[n_{j}]}}{\theta^{[N_{F}]}},
\]
where $x^{[j]}\equiv x(x+1)\cdots (x+j-1)$.
In the limit $N_{F}\to \infty$, the probability function
$P(\vec{n})$ becomes a Dirichlet distribution.
\[
P(\vec{x}\equiv \vec{n}/N_{F})=\lim_{N_{F}\to \infty}P(\vec{n})N_{F}^{2}
\frac{1}{\mbox{B}(\phi \vec{\mu})}
\prod_{k}x_{k}^{\mu_{k}\phi-1}.
\]

\bibliographystyle{cite}

\end{document}